\documentclass{iopart}
\begin{document}
\letter{A simple iterative algorithm  
for generating selected eigenspaces of large  matrices}
\author{F. Andreozzi, A. Porrino, and N. Lo Iudice}
\address{Dipartimento di Scienze Fisiche, Universit\`a di Napoli Federico 
II,\\
and Istituto Nazionale di Fisica Nucleare. \\
Complesso Universitario di Monte S. Angelo, Via Cintia, 80126 Napoli, Italy}
 
\begin{abstract}
We propose a new iterative algorithm for generating a subset
of eigenvalues and eigenvectors of large matrices which generalizes the
method of optimal relaxations. We also
give convergence criteria for the iterative process,   
investigate its efficiency by evaluating  
computer storage and time requirements and by few
numerical tests.
\end{abstract}

\maketitle
 
The increasing computational power has stimulated a growing interest toward 
developing and refining methods which allow to determine selected 
eigenvalues of a complex quantum system with extreme accuracy.
Widely adopted, specially for computing ground state properties,
are the quantum Monte Carlo methods 
\cite{MC}, where a properly defined function of the Hamiltonian is used as a
stochastic matrix which  guides a Markov process to sample the basis.

Alternatively, one may resort to direct diagonalization methods, like 
the Lanczos \cite{Lanc} and Davidson \cite{Dav} algorithms, of wide use
in several branches of physics.
The critical points of direct diagonalization methods are the amount of 
memory needed  and the time spent
in the diagonalization process. Because of these limitations, several 
systems are still out of reach even with the computer power now available. 
 
In this paper we present an iterative method, extremely easy to 
implement, for generating a subset of eigenvectors 
of a large matrix, give convergence criteria and show that it represents
a generalization of the method of optimal relaxations \cite{Shavitt}.

We assume first that the matrix $A$ represents
a self-adjoint operator $\hat A$ in an orthonormal basis 
$\{\mid 1 \rangle,\mid 2 \rangle,\dots,\mid N 
\rangle \}$ and is symmetric 
($a_{ij} = \langle i \mid \hat{A} \mid j \rangle = a_{ji}$).  
For the sake of simplicity, we illustrate the procedure for
a one-dimensional eigenspace.
The algorithm consists of a first approximation loop and subsequent 
iterations of refinement loops. 
The first loop goes  through the following steps:

{\bf 1a)} Start with the first two basis vectors and diagonalize 
the matrix
$\left( \begin{array}{cc} \lambda_1^{(1)} & a_{12} \\ a_{12} & a_{22}
\end{array} \right) $, where $\lambda_1^{(1)}=a_{11}$.

{\bf 1b)} Select the eigenvalue $\lambda_2^{(1)}$ and the
corresponding eigenvector $\mid \phi_2^{(1)} \rangle = 
K_{2,1}^{(1)} \mid \phi_1^{(1)} \rangle + K_{2,2}^{(1)} \mid 2 
\rangle$, 
where $\mid \phi_1^{(1)} \rangle \equiv \mid 1 \rangle$. 

\vskip0.1truecm

{\bf for} $j = 3, \dots, N$

\hskip0.5truecm {\bf 1c)} compute $b_j^{(1)} = 
\langle \phi_{j-1}^{(1)} \mid \hat A \mid j \rangle$.

\hskip0.5truecm {\bf 1d)} Diagonalize the matrix~~~~~~~ 
$\left( \begin{array}{cc} \lambda_{j-1}^{(1)} & b_j^{(1)} \\ b_{j}^{(1)} 
& a_{jj}
\end{array} \right) . $

\hskip0.5truecm {\bf 1e)} Select the eigenvalue $\lambda_j^{(1)}$ and the
corresponding eigenvector $\mid \phi_j^{(1)} \rangle$.

{\bf end} $j$

\vskip0.2truecm

\noindent The first loop yields an approximate eigenvalue
$\lambda_N^{(1)} \equiv E^{(1)} \equiv \lambda_0^{(2)}$ and an
approximate eigenvector $\mid \psi^{(1)} \rangle \equiv
\mid \phi_N^{(1)} \rangle \equiv \mid \phi_0^{(2)} \rangle =
\sum_{i=1}^N K_{N,i}^{(1)} \mid i \rangle$. With these new entries 
we start an iterative procedure which goes through the following  
refinement loops: 

{\bf for} $n = 2,3, \dots \;\;\;$ till convergence  

\hskip0.5truecm {\bf for} $j = 1,2, \dots, N$

\hskip1.0truecm {\bf 2a)} Compute $b_j^{(n)} = 
\langle \phi_{j-1}^{(n)} \mid \hat{A} \mid j \rangle$.

\hskip1.0truecm {\bf 2b)} Solve the eigenvalue problem in the   
general form
 $$det \; 
\left[ \;
\left ( \begin{array}{cc} \lambda_{j-1}^{(n)} & b_j^{(n)} \\ b_{j}^{(n)} 
& a_{jj} \end{array} \right) 
\; - \; \lambda \;
\left( \begin{array}{cc} 1 & K_{j-1,j}^{(n)} \\ K_{j-1,j}^{(n)} 
& 1 \end{array} \right) \; \right]
\; = \;0 $$
where the appearance of the metric matrix follows 
from the non orthogonality of the redifined basis $\mid 
\phi_{j-1}^{n}\rangle$ and $\mid j \rangle$.

\hskip1.0truecm {\bf 2c)} Select the eigenvalue $\lambda_j^{(n)}$ and the
corresponding eigenvector $\mid \phi_j^{(n)} \rangle$.

\hskip0.5truecm {\bf end} $j$

{\bf end} $n$.

\vskip0.2truecm

\noindent 
The $n$-th loop yields an approximate eigenvalue
$\lambda_N^{(n)} \equiv E^{(n)} \equiv \lambda_0^{(n+1)}$.
As for the eigenvector, at any step of the 
$j$-loop,  we have
\begin{equation}
\mid \phi_j^{(n)} \rangle \;=\; p_j^{(n)} \; \mid \phi_{j-1}^{(n)} \rangle \;+
\; q_j^{(n)} \;\mid j \rangle, \label{phi}
\end{equation}
with the appropriate normalization condition
$[p_j^{(n)}]^2\;+\;[q_j^{(n)}]^2\;+\;2\;p_j^{(n)}q_j^{(n)}K_{j-1,j}^{(n)}
\;=\;1$. The iteration of Eq. (\ref{phi}) yields the $n$-th eigenvector
\begin{equation}
\mid \psi^{(n)} \rangle  \equiv \mid \phi_N^{(n)} \rangle
\;=\; P_0^{(n)} \; \mid \psi^{(n-1)} \rangle \;+
\; \sum_{i=1}^{N} \;P_i^{(n)} \;q_i^{(n)} \;\mid i \rangle, \label{psi} 
\end{equation}
where the numbers $P_i^{(n)}$ are  
\begin{equation}
P_i^{(n)} \; = \; \prod_{k=i+1}^N \; p_k^{(n)} \;\;\; (i=0,1,\dots,N-1)
\;\;\; ; \;\; P_N^{(n)}\;=\;1. \nonumber
\end{equation}
The algorithm defines therefore the sequence of vectors (2),
whose convergence properties we can now examine. 
The  $q_j^{(n)}$
and $p_j^{(n)}$ coefficients can be expressed as
\begin{eqnarray}
q_j^{(n)}&\;=\;&{ {\mid B_j^{(n)} \mid} \over{
\Big[ (a_{jj}K_{j-1,j}^{(n)}-b_j^{(n)})^2 \;+ \; 2
K_{j-1,j}^{(n)}\;(a_{jj}K_{j-1,j}^{(n)}-b_j^{(n)})B_j^{(n)} \;+ \; 
(B_j^{(n)})^2 \Big]^{1 \over 2}}},\nonumber\\
p_j^{(n)} &\;=\;&(a_{jj}K_{j-1,j}^{(n)}-b_j^{(n)}) \;
{ q_j^{(n)} \over B_j^{(n)} } \label{pj},
\end{eqnarray}
where
\begin{equation}
B_j^{(n)} \;=\; 
\Big[ 
\lambda_{j-1}^{(n)}-\lambda_{j}^{(n)} 
\Big]
\;-\; K_{j-1,j}^{(n)} 
\Big[ 
(a_{jj}-\lambda_{j}^{(n)}) (\lambda_{j-1}^{(n)} - \lambda_j^{(n)})
{\Big]}^{1 \over 2}.
\end{equation}
It is apparent from these relations that, if
\begin{equation}
\mid \lambda_{j-1}^{(n)} - \lambda_j^{(n)} \mid \; \rightarrow \;0,
\;\;\;\; \forall j,\label{conv}
\end{equation}
the sequence 
$\mid \psi^{(n)} \rangle$ has a limit
$\mid \psi \rangle$, which is an eigenvector of the matrix $A$.
In fact, defining the residual vectors
\begin{equation}
\mid r^{(n)} \rangle  =  ( \hat{A} - E^{(n)} ) \; \mid \psi^{(n)} \rangle, 
\end{equation}
a direct computation gives for their components
\begin{eqnarray}
r_l^{(n)} &=& p_N^{(n)}  
\Big[ (a_{ll}-\lambda_{l}^{(n)}) (\lambda_{l-1}^{(n)} - \lambda_l^{(n)})
{\Big]}^{1 \over 2} 
+ q_N^{(n)} \; \Big\{ a_{lN} - \lambda_{N}^{(n)} \delta_{lN} \Big\}
\nonumber \\
&-&  p_N^{(n)} \Big\{ 
\big( \lambda_{l-1}^{(n)}-\lambda_{l}^{(n)} \big) 
\; K_{l,l-1}^{(n)}
+ \big( \lambda_{N-1}^{(n)}-\lambda_{N}^{(n)} 
\big) \; K_{l,N-1}^{(n)} \Big\}.
\end{eqnarray}
In virtue of (\ref{conv}), the norm of the $n$-th residual vector
converges to zero, namely $\mid\mid r^{(n)}\mid\mid
\rightarrow 0$. Equation (\ref{conv}) gives therefore a necessary condition 
for the convergence of  
$\mid \psi^{(n)}\rangle$ to an eigenvector
$\mid \psi\rangle$ of $A$,
with a corresponding eigenvalue $E=lim \; E^{(n)}$.
This condition holds independently of the prescription adopted for
selecting the eigensolution. Indeed,
we never had to specify the selection rule in steps  1b), 1e) and 2c).
Equation (\ref{conv}) is not only a necessary but also a sufficient condition 
for the convergence to the lowest or the highest eigenvalue of $A$.
In fact, the sequence $\lambda_j^{(n)}$ is monotonic (decreasing or increasing,
respectively), bounded from below or from above by the trace and therefore 
convergent.

The just outlined algorithm has a variational foundation. Its 
variational counterpart is just
the method of optimal relaxation \cite{Shavitt}. Indeed, for
the $p_j^{(n)}$ and $q_j^{(n)}$  given by Eqs. (\ref{pj}),
the $\alpha_j^{(n)}( = q_j^{(n)}/p_j^{(n)})$ derivative of the Rayleigh quotient
\begin{equation}
\rho(\phi_j^{(n)}) \;=\; 
{{\langle \phi_{j}^{(n)} \mid \hat{A} \mid \phi_{j}^{(n)} \rangle} \over 
{\langle \phi_{j}^{(n)} \mid \phi_{j}^{(n)} \rangle}}
\end{equation}
vanishes identically.

On the other hand the present matrix formulation allows in a 
straightforward way for  the  optimal
relaxation of an arbitrary number $t$ of coordinates, thereby 
improving the convergence rate of the procedure. 
We only need to turn the two-dimensional into a $(t+1)$-dimensional 
eigenvalue problem in steps 1d) and 2b),
compute t elements $b_j$ in steps 1c) and 2a), and accordingly redefine the
$j$-loops. The current eigenvector is still defined by the iterative
relation ($\alpha^{(n)}_{kj}= q^{(n)}_{k}/p^{(n)}_{j}$)
\begin{equation}
\mid \phi_{j+t}^{(n)} \rangle \;=\; p_j^{(n)} \left(\; \mid \phi_{j}^{(n)}
\rangle \; + \sum_{k=j+1}^{j+t}
\; \alpha_{kj}^{(n)} \;\mid k \rangle \right),
\end{equation}
which automatically fulfils the extremal conditions
\begin{equation}
{\partial \over {\partial \alpha_{kj}^{(n)}}} \rho(\phi_{j+t}^{(n)})=0, \; \; 
k=j+1,...,j+t. 
\end{equation}
Moreover, the algorithm can be naturally extended to generate 
at once an arbitrary number $m$ of lowest eigenstates.
We have simply to replace the two-dimensional matrices with
multidimensional ones having the following block structure: A $m \times m$ 
submatrix diagonal in the selected $m$ eigenvalues, which replaces 
$\lambda_{j-1}^{(n)}$, a $m' \times m'$ submatrix corresponding to $a_{jj}$ 
and two  $m \times m'$ off-diagonal blocks replacing
$b_j^{(n)}$ or $K_{j-1,j}^{(n)}$. 
This new formulation amounts to
an optimal relaxation method of several coordinates into a 
multidimensional subspace. It avoids therefore the use of
deflation or shift techniques for the computation of higher 
eigenvalues and eigenvectors.

It remains now to investigate the practical feasibility of the method.
The main issues to be faced are the storage and time requirements.
In the one-dimensional case, we need to store a single $N$-dimensional vector 
(the eigenvector).
The time is mainly determined by the $j$ loop. 
This requires $N$ operations for implementing point 2a) plus
$k \simeq 15$ remaining operations. Since $n= 1,2,...,n_c$ and 
$j=1,2,...,N$, the algorithm requires altogether $n_c (N^2 + kN)$
operations. 
It follows that, for large dimensional matrices, the number of
operations grows like $N^{2}$. For sparse matrices with  
an average number $L$ of non zero matrix elements, 
the required number of operations is $n_c (L + k)N$ and
therefore grows linearly with $N$. 
In the multidimensional case we need to store $m$ $N$-dimensional vectors.
If necessary, however, we can keep only one at a time and
store the remaining $m-1$ vectors in a secondary storage. 
This latter feature clearly shows 
that the algorithm lends itself to a straightforward parallelization. Also in 
the multidimensional case, the number of operations grows as $n_c m N^2$.

The algorithm has other remarkable properties: 
i) It works perfectly even in case of degeneracy of the eigenvalues. 
ii) The diagonalization
of the submatrices of order $m+m'$ insures the orthogonalization of the
full N-dimensional eigenvectors at each step. Therefore, 
no {\it ghost} eigenvalues occur.
iii) The range of validity of the algorithm can be easily enlarged if we remove
some of the initial assumptions. Clearly, the iterative 
procedure applies to a
non-orthogonal basis. We simply need to substitute steps 1a) and
1d) of the first loop with the appropriate generalized
eigenvalue problem. It applies also to non symmetric matrices. 
We have only to update both right and 
left eigenvectors and perform
steps 1c) and  2a)  for both non-diagonal matrix elements.
 
In order to test the efficiency and the convergence rate of the 
iterative procedure, we have applied the method to several examples.
The first is a 5-point finite difference matrix arising from the 
two-dimensional Laplace equation \cite{sparse}.
This is a block-tridiagonal 
matrix of $n_{b}$ $b$-dimensional blocks, whose eigenvalues are
\begin{equation}
\lambda_{ij} = 4\left(sin^{2}\frac{i\pi}{2(n_{b}+1)} + 
sin^{2}\frac{j\pi}{2(b+1)}\right),
\end{equation}
where $i=1,2\ldots, n_{b}$ and $j=1,2\ldots, b$. As in  
\cite{sparse}, we considered a block-matrix
with $n_{b}=15$ and $b=20$. We have tested the one-dimensional as well
as the multidimensional version of the algorithm. 
As shown in  Figure \ref{fig1}, the iterative procedure 
converges much faster in the multidimensional case. In fact,
the convergence rate increases with 
the number $\nu$ of generated eigenvalues and is
considerably faster than in Lanczos. It is also to be stressed 
that our algorithm
allows for an arbitrarily high accuracy, up to the machine 
precision limit.
The method, specially in its multidimensional extension, is quite 
effective even if applied to the same matrix  with $n_{b}=b$ so as 
to allow for degeneracy. For $n_{b}=b=80$, it yields the lowest 
seven roots, including two couples of degenerate eigenvalues, 
with an accuracy of $10^{-12}$.
 
A second example, still taken from \cite{sparse}, 
is a one-dimensional biharmonic band matrix 
whose eigenvalues  
\begin{equation}
\lambda_{k} = 16 sin^{4}\frac{k\pi}{2(N+1)}\; \; \; \; \; k = 
1,\ldots, N 
\end{equation}
are small  
and poorly separated from each other. A similarly high density 
of levels occurs in the Anderson 
model of localization \cite{Elsn}. Because of this peculiarity, 
the limit of the machine precision is reached for a modest increase
of the dimension $N$ of the matrix. Our method reproduces perfectly 
any number of eigenvalues with the required accuracy after a   
small number of iterations. In the specific example discussed in   
\cite{sparse} ($N=20$) we attained the highest accuracy after eight iterations,
much less than all methods discussed there. We have checked that, 
unlike others, our method works without any modification even if we 
increase the dimension $N$ up to the limit compatible with the machine precision.
In this case the number of iterations needed increase by an order of 
magnitude, in any case, below 100.

A third example is provided by a matrix with diagonal matrix elements $ 
a_{ii} = 2 \sqrt{i} - a$ and off-diagonal ones $a_{ij} =-a$
or $a_{ij} = 0$ according that they fall within or outside
a band of width $2L$. Such a matrix simulates a pairing Hamiltonian
relevant to many branches of physics.
We have considered a matrix of dimension $N= 10^{8}$ and half-band width 
$L=400$. We found convergence after $28$ iterations, reaching an 
accuracy of $10^{-8}$ for the eigenvalues. The time required to compute 
the lowest eigenvalue through  the 
one-dimensional algorithm is $t=18697$ s for 
a workstation of $500$ MHz and $512$ Mb of RAM.  

Finally, we generalize the latter example by considering
a full matrix of dimension $N=10^{5}$ with 
matrix elements $ a_{ij}= 2 \sqrt{i} \delta_{ij} + (-1)^{i+j} a 
\frac{i+j}{\sqrt{i^{2} + j^{2}}}$. Their alternating signs are also to be 
noticed, since they decrease somewhat the rate of convergence of the process.
We reproduce the lowest eigenvalue with  
an accuracy of $10^{-5},10^{-6},10^{-7},10^{-8}$ 
after $n_{c}=42,70,155,330$ iterations respectively.
 
In conclusion, the present diagonalization algorithm is a 
generalization of the variational optimal relaxation method and,
on the ground of the examples discussed, appears
to be more competitive than the 
methods currently adopted. 
It seems to be faster 
and to require a minimal amount of computer storage.
It is extremely simple to be implemented and is {\it robust}, yielding
always stable numerical solutions. Moreover, it is free of 
{\it  ghost} eigenvalues.  
Because of these features, we are confident that it can be applied 
quite effectively to 
physical systems, like medium-light nuclei or quantum dots with few 
electrons.  

\bigskip
\ack{The work was partly supported by the Prin 99 of the Italian MURST}

\Figures

\begin{figure}
\caption{\label{fig1}Convergence rate of the present algorithm 
applied to a finite difference matrix deduced from Laplace equation 
\cite{sparse} for different numbers $\nu$ of generated eigenvalues.
The data referring to the Lanczos convergenge rate are taken from 
\cite{sparse}}
\end{figure}

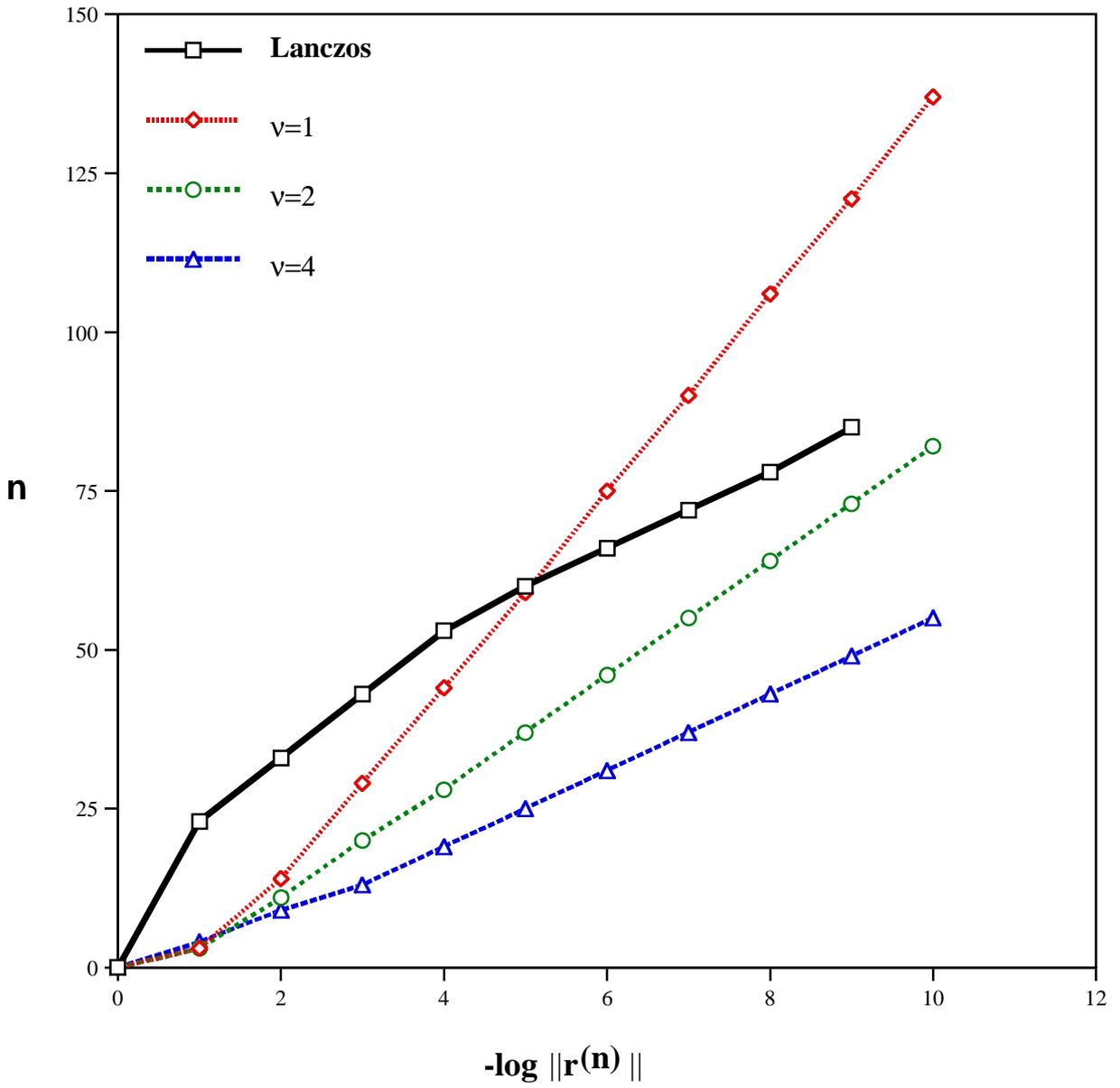
\end{document}